\documentclass[aps,prb,twocolumn,floats,showpacs]{revtex4}
\usepackage{graphicx}
\usepackage{amsmath}
\def\stacksymbols #1#2#3#4{\def\theguybelow{#2}
    \def\verticalposition{\lower#3pt}
    \def\spacingwithinsymbol{\baselineskip0pt\lineskip#4pt}
    \mathrel{\mathpalette\intermediary#1}}
\def\intermediary#1#2{\verticalposition\vbox{\spacingwithinsymbol
      \everycr={}\tabskip0pt
      \halign{$\mathsurround0pt#1\hfil##\hfil$\crcr#2\crcr
               \theguybelow\crcr}}}

\begin{document}
\title{Anomalously localized states and multifractal correlation of critical wavefunctions\\
in two-dimensional electron systems with spin-orbital interactions
}

\author{H. Obuse and K. Yakubo}

\affiliation{Department of Applied Physics, Hokkaido University, Sapporo 060-8628, Japan.}

\begin{abstract}
Anomalously localized states (ALS) at the critical point of the
Anderson transition are studied for the SU(2) model belonging to
the two-dimensional symplectic class. Giving a quantitative
definition of ALS to clarify statistical properties of them, the
system-size dependence of a probability to find ALS at criticality
is presented. It is found that the probability {\it increases}
with the system size and ALS exist with a finite probability even
in an infinite critical system, though the typical critical states
are kept to be multifractal. This fact implies that ALS should be
eliminated from an ensemble of critical states when studying
critical properties from distributions of critical quantities. As
a demonstration of the effect of ALS to critical properties, we
show that the distribution function of the correlation dimension
$D_2$ of critical wavefunctions becomes a delta function in the
thermodynamic limit only if ALS are eliminated.
\end{abstract}


\maketitle

\section{Introduction}
\label{sec:1}

The disorder induced metal-insulator transition, namely, the
Anderson transition,\cite{Janssen3} corresponds to a fixed point
in a real-space renormalization transformation. Approaching the
critical point from an insulating phase, the localization length
$\xi$ increases and diverges at the critical point. Since the
length $\xi$ is a unique characteristic length near the transition
point, there is no length scale characterizing the critical state.
This implies that the wavefunction at the Anderson transition
point (the critical wavefunction) is scale invariant. In fact,
Aoki \cite{Aoki1} and Wegner \cite{Wegner1} have demonstrated that
critical wavefunctions have multifractal properties which can be
described by an infinite set of exponents. These exponents are
quite important because they define not only spatial distribution
of wavefunction amplitudes but also dynamical properties of the
electron system at criticality.\cite{Chalker1}

Among various works in which exponents characterizing
multifractality of critical wavefunctions have been extensively
studied,
\cite{Chalker1,Schreiber1,Pook1,Ketzmerick1,Janssen2,Huckestein1,Grussbach1,Mirlin2,Evers1,Nakayama2}
numerical multifractal analyses based on the box-counting method
give the most reliable values of exponents at the present stage,
except for a few exactly-solvable problems. \cite{Castillo1}
However, values of exponents reported so far widely fluctuate. For
example, the correlation dimension $D_2$ for the three-dimensional
Anderson model takes values ranging from $1.45$ \cite{Grussbach1}
to $1.68$.\cite{Schreiber1} These fluctuations result mainly from
the following reasons. One is a finite-size effect which can be
excluded in principle by a finite-size scaling. Another reason is
related to {\it anomalously localized states}
(ALS)\cite{Al'tshuler1,Mirlin1} at the critical point. Statistical
properties of ALS in a metallic phase have been studied well.
\cite{Muzykantskii1,Falko1,Smolyarenko1,Uski1,Nikolic1,Nikolic2,Uski2}
Specific local configurations of disorder in a finite spatial
region may lead to electron states localized within this region,
while the average disorder is weak compared to the critical
disorder. Although such ALS induced by statistical fluctuations of
disorder configurations remain even if the system is infinite,
they do not contribute to ensemble-averaged properties of
amplitude distributions of wavefunctions or transport properties
in the metallic phase. This is because ALS have point-like
spectra, while metallic (extended) states have continuous ones.

ALS also appear at the critical
point.\cite{Mirlin2,Canali1,Steiner1} The localization length of a
multifractal wavefunction diverges in an infinite system at the
critical point, while sizes of ALS are finite. In contrast to the
case of metallic phases, it is not obvious that ALS can be
neglected in infinite systems at the critical point. The effect of
ALS is much stronger than that in metallic phases.
However, fundamental knowledge on, e.g., a probability to find ALS
or influences of ALS to critical properties, are still unclear
because of the lack of a quantitative definition of ALS. It is
important to clarify how strong ALS affect critical properties by
defining ALS quantitatively and revealing their fundamental
features.

In the present paper, we give a quantitative and practical
definition of ALS at criticality suitable for numerical
investigations. In order to define ALS, it is natural to evaluate
spatial extend of wavefunctions. Quantum states with localization
lengths less than the system size are regarded as ALS in a finite
system. This definition is, however, not appropriate for precise
and systematic calculations, because we cannot determine which
wavefunctions are more multifractal if localization lengths of two
wavefunctions are both close to the system size. We show that the
correlation function of box-measures of multifractal critical
wavefunctions \cite{Janssen1,Pracz1,Janssen2} works quite well for
the definition of ALS. In order to demonstrate how ALS influence
critical properties of the Anderson transition, we examine the
distribution function of the correlation dimension $D_2$ of
critical wavefunctions. Results presented here urge us to
reconsider our picture of ALS.

The efficiency of our definition of ALS is demonstrated by
applying them to critical wavefunctions in two-dimensional
electron systems with spin-orbit interactions (symplectic
systems), which exhibit the Anderson metal-insulator transition.
For ensuring the multifractality of typical critical wavefunctions
even in a short-range scale, we adopt the SU(2) model
\cite{Asada1} in which scaling corrections due to irrelevant
scaling variables are negligible.

This paper is organized as follows. In Sec.~\ref{sec:2}, we
propose a quantitative definition of ALS based on the correlation
function of box-measures of wavefunction amplitudes. Some quantities and exponents
appearing in the multifractal analysis are defined in this
section. In Sec.~\ref{sec:3}, we briefly explain the SU(2) model
for which the efficiency of our methods is demonstrated and glance
a numerical technique to obtain eigenstates of the SU(2) model.
Several ALS calculated numerically for the SU(2) model are shown
in Sec.~\ref{sec:4}. We display how well our definition of ALS
works via concrete examples of ALS in the SU(2) model.
In Sec.~\ref{sec:5},
we investigate the fluctuation of values of the correlation dimension $D_2$ in infinite SU(2) systems
as an example of the effect of ALS.
Section~\ref{sec:6} is devoted to conclusions.

\section{Definition of Anomalously Localized States}
\label{sec:2}

In this section, we propose a quantitative definition of ALS based
on a multifractal analysis. For this purpose, we give some
definitions of basic quantities and exponents used in the
multifractal analysis. At first, we introduce a quantity $Z(q)$
defined by
\begin{equation}
Z_q (l) = \sum_b \mu_{b (l)}^q,
\label{eq:1}
\end{equation}
where $\mu_{b(l)}=\sum_{i \in b (l)} |\psi_i|^2$ [or
$\mu_{b(l)}=\int_{b(l)} |\psi(\boldsymbol{r})|^2 d \boldsymbol{r}$
in continuous systems] is the box measure of wavefunction
amplitudes. The summation $\sum_{i \in b(l)}$ (or the integral
$\int_{b(l)}d \boldsymbol{r}$) is taken over sites $i$ (or the
spatial region) within a small box $b$ of size $l$, and the
summation in Eq.~(\ref{eq:1}) is taken over such boxes. The
quantity $q$ is the order of the moment $Z_q(l)$. For a
multifractal wavefunction, $Z_{q}(l)$ obeys a power law
\begin{equation}
Z_q (l) \propto l^{\tau(q)},
\label{eq:2}
\end{equation}
where $\tau(q)$ is called the mass exponent.
From Eq.~(\ref{eq:2}), the mass exponent is given by
\begin{equation}
\tau(q) = \lim_{l \rightarrow 0} \frac{\ln Z_q(l)}{\ln l}.
\label{eq:3}
\end{equation}

Since ALS are not multifractal, it seems possible to define ALS by
calculating deviations of the moment $Z_q(l)$ from the power law
Eq.~(\ref{eq:2}). However, the moment $Z_q(l)$ represents simply
the average of $\mu_{b(l)}^q$ [except for a trifling prefactor
$(L/l)^d$], and does not describe the multifractal correlation of
box measures. Thus, the quantity $Z_q(l)$ is not sensitive to ALS
and Eq.~(\ref{eq:2}) is insufficient for the definition of ALS.
This will be demonstrated by showing an example in
Sec.~\ref{sec:4}.

The correlation function $G_q (l,L,r)$ defined below can describe
the box-measure correlations:\cite{Cates1,Janssen2}
\begin{equation}
G_q (l,L,r) = \frac{1}{N_b N_{b_r}} \sum_b \sum_{b_r} \mu_{b(l)}^q \mu_{b_{r}(l)}^q ,
\label{eq:4}
\end{equation}
where $\mu_{b_{r}(l)}$ is the box measure of a box $b_r(l)$ of
size $l$ fixed distance $r-l$ away from the box $b(l)$, $N_b$ (or
$N_{b_r}$) is the number of boxes $b(l)$ [or $b_r(l)$], and the
summation $\sum_{b_r}$ is taken over all such boxes $b_r(l)$. The
correlation function $G_q(l,L,r)$ is a generalized quantity of the
$q$th moment $Z_q(l)$. For a multifractal wavefunction,
$G_q(l,L,r)$ should behave as \cite{Janssen2}
\begin{equation}
G_q(l,L,r) \propto l^{x(q)} L^{-y(q)} r^{-z(q)},
\label{eq:5}
\end{equation}
where $x(q)$, $y(q)$, and $z(q)$ are exponents in multifractal correlations.
This proportionality relation is sensitive to ALS and then
suitable for defining ALS. Since our purpose is to examine
multifractality of individual wavefunctions by using
Eq.~(\ref{eq:5}), the system size $L$ is always fixed. To find the
$l$ and $r$ dependences of $G_q(l,L,r)$, we concentrate on the
following functions,
\begin{equation}
Q_L(l) = G_2(l,L,r=l) \propto l^{x(2)-z(2)},
\label{eq:6}
\end{equation}
and
\begin{equation}
R_L(r) = G_2(l=1,L,r) \propto r^{-z(2)}.
\label{eq:7}
\end{equation}
In order to quantify non-multifractality of a specific
wavefunctions, it is convenient to introduce variances
$\text{Var} (\log_{10} Q_L)$ and $\text{Var} (\log_{10} R_L)$ from the
linear functions of $\log_{10} l$ and $\log_{10} r$, $\log_{10} Q_{L}(l) =
[x(2)-z(2)] \log_{10} l + c_Q$ and $\log_{10} R_{L}(r) = -z(2) \log_{10} r +
c_R$, respectively, calculated by the least-square fit. From these
variances, a quantity $\Gamma$ is defined by
\begin{equation}
\Gamma(L,\lambda) = \lambda \text{Var}(\log_{10} Q_L) + \text{Var} (\log_{10} R_L),
\label{eq:8}
\end{equation}
where $\lambda$ is a factor to compensate the difference between
average values of $\text{Var}(\log_{10} Q_L)$ and $\text{Var}(\log_{10}
R_L)$. Using $\Gamma$ given by Eq.~(\ref{eq:8}), the quantitative
and expediential definition of ALS at criticality is presented by
\begin{equation}
\Gamma > \Gamma^*,
\label{eq:9}
\end{equation}
where $\Gamma^*$ is a criterial value of $\Gamma$ to distinguish
ALS from multifractal states and chosen appropriately as
demonstrated later.

\section{Model}
\label{sec:3}

In this paper, the efficiency of the above definition of ALS and
the effect of ALS will be demonstrated for an ensemble of
wavefunctions at the Anderson metal-insulator transition. For this
purpose, the effect of scaling corrections due to irrelevant
fields which is a finite-size effect and independent of the ALS
should be reduced as much as possible. Therefore, considering the
advantage of system sizes, we focus our attention on the Anderson
transition in two-dimensional electron systems with strong
spin-orbit interactions, in which systems have no spin-rotational
symmetry but have the time-reversal one. Hamiltonians describing
these systems belong to the symplectic ensemble. Several models
\cite{Evangelou1,Ando1,Merkt1} have been proposed so far to
represent the symplectic systems. Although the Ando model
\cite{Ando1} has been most extensively studied, relatively large
scaling corrections make it difficult to distinguish the ALS
effect. Recently, Asada {\it et al.}\cite{Asada1} proposed the
SU(2) model belonging to the symplectic class, for which scaling
corrections are negligibly small. While the SU(2) model is rather
mathematical, we believe that presented results are qualitatively
the same with those for other systems exhibiting the Anderson
transition.

The Hamiltonian of the SU(2) model is compactly written in a
quaternion representation as
\begin{equation}
\boldsymbol{H} = \sum_{i} \varepsilon_i  \boldsymbol{c}_{i}^\dagger  \boldsymbol{c}_{i}
  -  V \sum_{i,j} \boldsymbol{R}_{ij} \boldsymbol{c}_{i}^\dagger \boldsymbol{c}_{j},
\label{eq:10}
\end{equation}
where $ \boldsymbol{c}_{i}^\dagger$ ($ \boldsymbol{c}_{i}$) is the
creation (annihilation) operator acting on a quaternion state
vector, $\boldsymbol{R}_{ij}$ is the quaternion-real hopping
matrix element between the sites $i$ and $j$, and $\varepsilon_i$
denotes the on-site random potential distributed uniformly in the
interval $[-W/2, W/2]$. The strength of the hopping $V$ is taken
to be the unit of energy. Hereafter, we denote quaternion-real
quantities by bold symbols. A quaternion-real number
$\boldsymbol{x}$ can be written in the form
\begin{equation}
\boldsymbol{x} = \sum_{\mu=0}^{3} x_{\mu} \boldsymbol{\tau}^{\mu},
\label{eq:11}
\end{equation}
where $x_\mu$ is a real number and the primitive elements
$\boldsymbol{\tau}^\mu (\mu=0,1,2,3)$ define the quaternion
algebra as
\begin{eqnarray}
\boldsymbol{\tau}^0 \boldsymbol{\tau}^0 &=&
 -\boldsymbol{\tau}^n \boldsymbol{\tau}^n = \boldsymbol{\tau}^0  \quad (n=1,2,3), \\
\label{eq:12}
\boldsymbol{\tau}^0 \boldsymbol{\tau}^n &=&
 \boldsymbol{\tau}^n \boldsymbol{\tau}^0 = \boldsymbol{\tau}^n \quad (n=1,2,3),\\
\label{eq:13}
\boldsymbol{\tau}^l \boldsymbol{\tau}^m &=&
 -\boldsymbol{\tau}^m \boldsymbol{\tau}^l = \boldsymbol{\tau}^n
\label{eq:14} \\
&&(l,m,n=1,2,3 \ \text{and any cyclic permutation}). \nonumber
\end{eqnarray}
In the Hamiltonian Eq.~(\ref{eq:10}), the matrix element
$\boldsymbol{R}_{ij}$ is defined by
\begin{eqnarray}
\boldsymbol{R}_{ij} &=& \cos \alpha_{ij} \cos \beta_{ij} \boldsymbol{\tau}^0
+ \sin \gamma_{ij} \sin \beta_{ij}\boldsymbol{\tau}^1
\nonumber
\\
&-& \cos \gamma_{ij} \sin \beta_{ij}\boldsymbol{\tau}^2
+ \sin \alpha_{ij} \cos \beta_{ij}\boldsymbol{\tau}^3,
\label{eq:15}
\end{eqnarray}
where $\alpha_{ij}$ and $\gamma_{ij}$ are distributed uniformly in
the range of $[0,2\pi)$, and $\beta_{ij}$ is distributed according
to the probability density $P(\beta) d \beta = \sin(2\beta) d
\beta$ for $0 \le \beta \le \pi/2$.

\begin{figure*}[t]
\includegraphics[width=17cm]{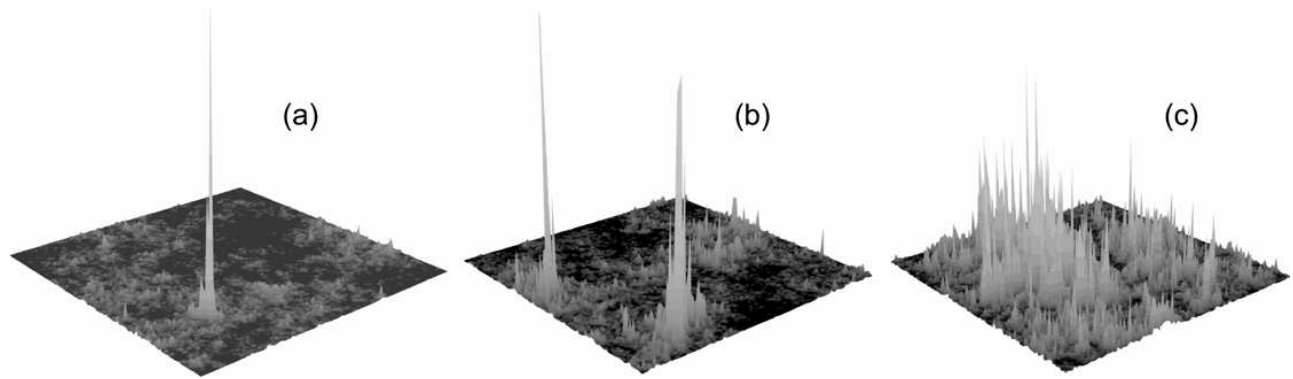}
\caption{ Squared amplitudes of critical wavefunctions
$|\psi_i|^2$ for the SU(2) model of the system size $L=120$. The
eigenenergies of these states are (a) $E=0.99976$, (b) $1.00020$,
and (c) $0.99970$. Wavefunctions labelled by (a) and (b) are
spatially localized and regarded as ALS. The wavefunction (c)
seems to be multifractal.} \label{fig:1}
\end{figure*}
\begin{figure*}[t]
\includegraphics[width=17.5cm]{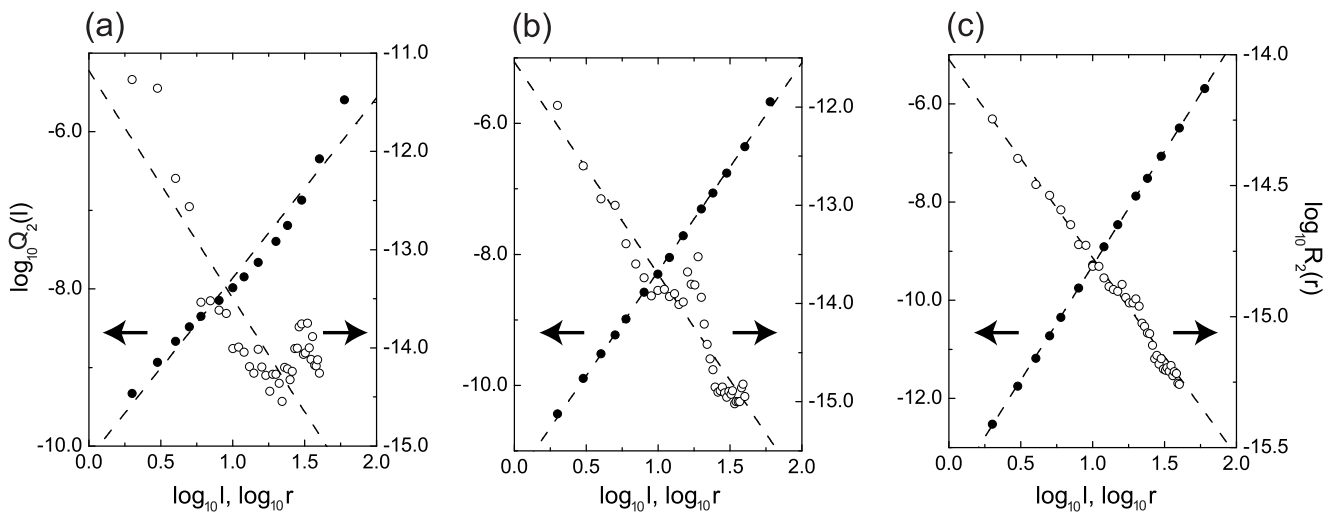}
\caption{
The functions $Q_L(l)$ (filled circles with the left vertical axis) and
$R_L(r)$ (open circles with the right axis) for the wavefunctions shown
in Fig.~1. The labels (a), (b), and (c) correspond to Figs.~1(a), 1(b), and 1(c),
respectively. Dashed lines are just guide to the eye.}
\label{fig:2}
\end{figure*}

We have calculated eigenvectors of the Hamiltonian
Eq.~(\ref{eq:10}) by using the forced oscillator method
(FOM).\cite{Nakayama1} The FOM is an efficient algorism to solve
quickly an eigenvalue problem with saving memory spaces, in which
an eigenvector belonging to the eigenenergy $E$ is extracted as a
resonant mode of the classical dynamical system described by the
same Hamiltonian to an external periodic force with the angular
frequency $\sqrt E$. The FOM can be extended to the eigenvalue
problem of quaternion-real matrices. The corresponding dynamical
system is described by
\begin{equation}
\frac{d^2}{dt^2} \boldsymbol{x}_i(t) =
- \sum_j \boldsymbol{H}_{ij} \boldsymbol{x}_j (t) + \boldsymbol{F}_i \cos (\Omega t),
\label{eq:16}
\end{equation}
where $\boldsymbol{H}_{ij}$ is the quaternion-real matrix
element of the Hamiltonian Eq.~(\ref{eq:10}),
$\boldsymbol{x}_i(t)$ is the displacement vector, and
$\boldsymbol{F}_i$ is the site-dependent amplitude of a
quaternion-real external force. Decomposing the displacement
$\boldsymbol{x}_i(t)$ into quaternion normal modes
$\boldsymbol{e}_i(\lambda)$ as $\boldsymbol{x}_i(t)
=\sum_\lambda \boldsymbol{e}_i(\lambda)
\boldsymbol{Q}_\lambda(t)$ , the quaternion coefficient
$\boldsymbol{Q}_\lambda(t)$ with $\lambda$ for which
$E_\lambda$ is the closest eigenenergy to $\Omega^2$ enlarges
compared to other coefficients after a long time. We should note
the order of normal modes $\boldsymbol{e}_i(\lambda)$ and
$\boldsymbol{Q}_\lambda(t)$ in the decomposition of
$\boldsymbol{x}_i(t)$ reflecting the non-commutative property
of quaternion numbers. An iterative procedure can accelerate the
enhancement of $\boldsymbol{Q}_\lambda(t)$. \cite{Obuse1}
Finally, we can obtain the eigenvector of the quaternion-real
matrix $\boldsymbol{H}$ as $\boldsymbol{x}_i(t) \sim
\boldsymbol{e}_i(\lambda)$. The obtained quaternion-real
eigenvector represents two physical states simultaneously, which
correspond to the Kramers doublet belonging to $E_\lambda$.
Denoting $\boldsymbol{e}_i(\lambda)$ by $\sum_{\mu}
e_i^\mu(\lambda) \boldsymbol{\tau}^\mu$, the $i$th elements
of the two complex vectors $|f(\lambda)\rangle$ and
$|g(\lambda)\rangle$ corresponding to the Kramers doublet are
given by $f_{2i-1}(\lambda)=e_i^0(\lambda)+ie_i^1(\lambda)$,
$f_{2i}(\lambda)=-e_i^2(\lambda)+ie_i^3(\lambda)$,
$g_{2i-1}(\lambda)=e_i^2(\lambda)+ie_i^3(\lambda)$, and
$g_{2i}(\lambda)=e_i^0(\lambda)-ie_i^1(\lambda)$,
respectively.\cite{Obuse1} As we expected, $|g(\lambda)\rangle$ is
the time-reversal vector of $|f(\lambda)\rangle$, namely,
$|g(\lambda)\rangle=-i \sigma_y |f(\lambda)\rangle$, where
$\sigma_y$ is the $y$ component of the Pauli matrix.

\section{Anomalously localized states for The SU(2) model at the critical point}
\label{sec:4}

In this section, we demonstrate by showing critical wavefunctions
of the SU(2) model that ALS can be discriminated by examining the
box-measure correlation function defined by Eq.~(\ref{eq:4}).
Critical wavefunctions are calculated by the FOM for the SU(2)
model with $W=5.952$ for which the eigenstate with $E=1$ is known
to be critical.\cite{Asada1} All wavefunctions calculated in this
paper have their eigenenergies closest to $E=1$. This implies that
we extract only one critical state from a system. In numerical
calculations, the periodic boundary conditions are employed for
both directions. Figure~\ref{fig:1} shows squared amplitudes of
critical wavefunctions in systems of $L=120$ with different
realizations of on-site random potentials. The wavefunctions shown
in Fig.~\ref{fig:1}(a) seems to be strongly localized with a
single peak of large amplitudes even for the critical conditions
($W=5.952$ and $E \approx 1$). We can regard this eigenstate as an
ALS. In order to exclude a possibility that the critical energy is
shifted due to finite-size effects, it has been confirmed that
eigenstates belonging to adjacent energy levels are not localized.
The wavefunction shown in Fig.~\ref{fig:1}(b) is also localized
with a relatively long localization length. In this case, we see
two peaks of large amplitudes in the spatial distribution of
squared amplitudes. On the contrary, the wavefunction shown in
Fig.~\ref{fig:1}(c) seems to be multifractal. The existence of
localized wavefunctions at criticality are not peculiar to the
SU(2) model. We have obtained localized wavefunctions also for the
Ando model at criticality.

In order to distinguish systematically ALS from wavefunctions
shown in Fig.~\ref{fig:1} by the box-measure correlation function,
we calculate $Q_L(l)$ and $R_L(r)$ for these three wavefunctions.
Results are shown in Fig.~\ref{fig:2} by filled $[ Q_L(l) ]$ and
open circles $[R_L(r)]$. Figures~\ref{fig:2}(a), \ref{fig:2}(b),
and \ref{fig:2}(c) correspond to the wavefunctions in
Figs.~\ref{fig:1}(a), \ref{fig:1}(b), and \ref{fig:1}(c),
respectively. Dashed lines are guide to the eye. The quantity
$Q_L(l)$ gives the averaged value over different ways to divide
the system into small boxes of size $l$. It is found that $Q_L(l)$
and $R_L(r)$ for the wavefunction Fig.~\ref{fig:1}(a) do not obey
power laws as shown in Fig.~\ref{fig:2}(a), which implies that the
wavefunction is not multifractal. The function $R_L(r)$ shown in
Fig.~\ref{fig:2}(b) does not follow a power law, while $Q_L(l)$
seems to be fit to a straight line. This is because the
wavefunction shown in Fig.~\ref{fig:1}(b) has two peaks of large
amplitudes and bears resemblance to the multifractal wavefunction
[Fig.~\ref{fig:1}(c)] in a local view. There exists, however, a
distinct characteristic length scale, namely, a distance between
two peaks. This fact prevents $R_L(r)$ from obeying a power law.
Since $Q_L(l)$ is the same with $N_b^{-1} Z_4(l)$,
Fig.~\ref{fig:2}(b) shows that $Z_q(l)$ is not sensitive to ALS.
On the contrary, both $Q_L(l)$ and $R_L(r)$ depicted in
Fig.~\ref{fig:2}(c) well follow power laws. This means that the
wavefunction shown in Fig.~\ref{fig:1}(c) is multifractal as we
expected.

\begin{figure}[t]
\includegraphics[width=8.5cm]{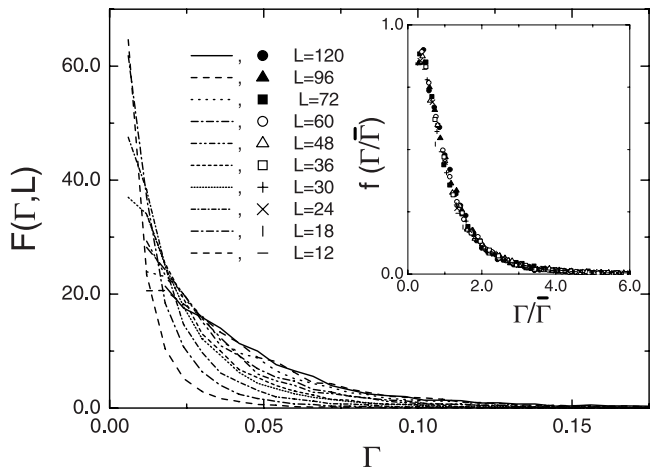}
\caption{ Distribution functions $F(\Gamma,L)$ versus $\Gamma$ for
system sizes $L=12$, $18$, $24$, $30$, $36$, $48$, $60$, $72$,
$96$, and $120$, where $\Gamma$ defined by Eq.~(\ref{eq:8}) with
$\lambda=3$ represents the degree of non-multifractality of a
wavefunction. The inset shows that the distribution function
$F(\Gamma,L)$ is characterized only by the average of $\Gamma$.}
\label{fig:3}
\end{figure}

We calculate the distribution function of the quantity $\Gamma$
defined by Eq.~(\ref{eq:8}) for ensembles of critical
wavefunctions. We prepare ten ensembles for different system sizes
($L=12$ to $120$). Each ensemble contains $10^4$ critical
wavefunctions. In the definition of $\Gamma$ [Eq.~(\ref{eq:8})],
we choose $\lambda=3$, because the average of $\text{Var} (\log_{10}
R_L)$ is about three times larger than that of $\text{Var} (\log_{10}
Q_L)$. Figure \ref{fig:3} shows the distribution functions
$F(\Gamma,L)$ versus $\Gamma$ for various system sizes $L$. Recall
that eigenstates having large values of $\Gamma$ are regarded as
ALS. As depicted in Fig.~3, the function $F(\Gamma,L)$ has a peak
at $\Gamma=0$ for any $L$. This gives an evidence that {\it
typical} critical states are multifractal. A remarkable feature is
that the function $F(\Gamma,L)$ becomes broad as the system size
increases. This implies that the probability to find ALS increases
with $L$, which obliges us to reconsider the role of ALS because
the influence of ALS to critical properties in infinite systems is
stronger than that in finite systems. These results lead to
crucial conclusions as follows: When we study critical properties
via the average value of quantities $X$ defined near the critical
point, the average must be the geometric mean or the mode value of
$X$, otherwise ALS disturb correct information on critical
properties. This condition has been satisfied by most of previous
works. Furthermore, when studying critical properties via
distributions of critical quantities such as the level statistics,
ALS should be eliminated from an ensemble of critical
wavefunctions to obtain precise information on criticality.

\begin{figure}[t]
\includegraphics[width=8cm]{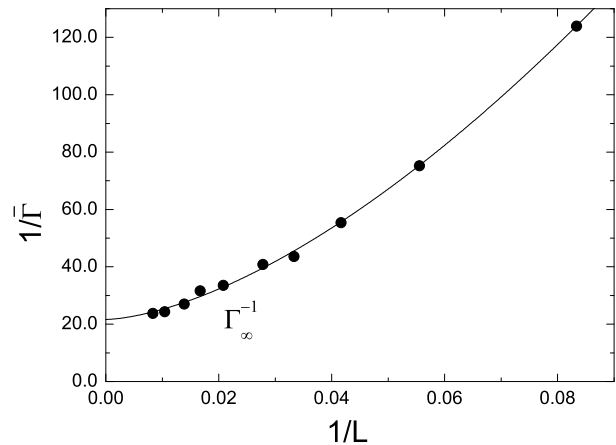}
\caption{System-size dependence of $\bar \Gamma$. Solid line
indicates a fit by $\bar \Gamma \ / \Gamma_\infty=1-c
L^{-\eta}$, where $\Gamma_\infty$ is the value of $\Gamma$ at $L
\rightarrow \infty$.} \label{fig:4}
\end{figure}

We also found that the distribution function $F(\Gamma,L)$ is
characterized only by the (arithmetic) average value of $\Gamma$
as shown in the inset of Fig.~3. Thus, the function $F(\Gamma,L)$
can be written in the form:
\begin{equation}
F \left( \Gamma,L \right) =
F_0 \left( \bar \Gamma\right)
f(\Gamma/\bar \Gamma)
\label{eq:17}
\end{equation}
where $\bar \Gamma$ is the average value of $\Gamma$ and is a
function of the system size $L$, $f$ is a function independent of
$L$, and $F_0$ is a normalization factor. The $L$ dependence of
$\bar \Gamma$ is shown in Fig.~4. We see from Fig.~4 that $\bar
\Gamma$ converges to a finite value $\Gamma_\infty$ for $L
\rightarrow \infty$, which implies that the probability to find
ALS in an infinite system remains to be finite. The fact that
$F(\Gamma,L)$ can be scaled only by $\bar \Gamma$ would be an
important statistical property of ALS. These results were obtained
for the SU(2) model. However, we believe that similar qualitative
features will be obtained for other systems at the Anderson
transition point.

\section{Fluctuations of the Correlation Dimension}
\label{sec:5}

In order to demonstrate how ALS affect critical properties in
infinite systems, we calculate the distribution of the correlation
dimension $D_2$ of critical wavefunctions of the SU(2) model. The
correlation dimension $D_2$ characterizes multifractality of
critical wavefunctions and is known to be related to some
exponents describing dynamical properties at the critical
point.\cite{Chalker1} It had been widely accepted that the
exponent $D_2$ is universal and does not depend on specific
samples of critical wavefunctions. The numerical work of Ref.~35,
however, claimed that the distribution function of $D_2$ has a
finite width even in the thermodynamic limit. In response to this
surprising result, recently, several
authors\cite{Mirlin2,Cuevas1,Varga1} have verified universality of
the exponent $D_2$ by precise and large-scale numerical
calculations. Their results exhibit that $D_2$ takes a definite
value independent of samples for infinite systems, which
contradicts the result by Ref.~35. In the following, we show that
the non-universal property of $D_2$ is a consequence of the
existence of ALS.

The correlation dimension $D_2$ is given by
\begin{equation}
Z_2(l) \propto l^{D_2},
\label{eq:18}
\end{equation}
where $Z_2(l)$ is defined by Eq.~(\ref{eq:1}), thus $D_2=\tau(2)$.
Although Eq.~(\ref{eq:18}) does not hold for ALS, we calculate
$D_2$ even for ALS by force by the least-square fit. The
distribution functions of $D_2$ for the same ensembles as for
Fig.~3 are shown in Fig.~5. The correlation dimension fluctuates
over samples in finite systems whereas the width of the
distribution function $P(D_2)$ becomes narrow as the system size
increases. This fluctuation results from the existence of ALS and
some scaling corrections due to a finite system-size. The
fluctuation of $D_2$ due to ALS remains even in infinite systems.
In order to obtain the distribution function of $D_2$ for typical
critical wavefunctions, we eliminate ALS defined by $\Gamma >
\Gamma^*$ from the original ensembles. To this end, we choose the
value of $\Gamma^*$ appeared in the definition of ALS
[Eq.~(\ref{eq:9})] as $\Gamma^*=0.03$. We study the system-size
dependence of the standard deviation $\sigma(L)$ of the
distribution function $P(D_2)$.

\begin{figure}[t]
\includegraphics[width=8cm]{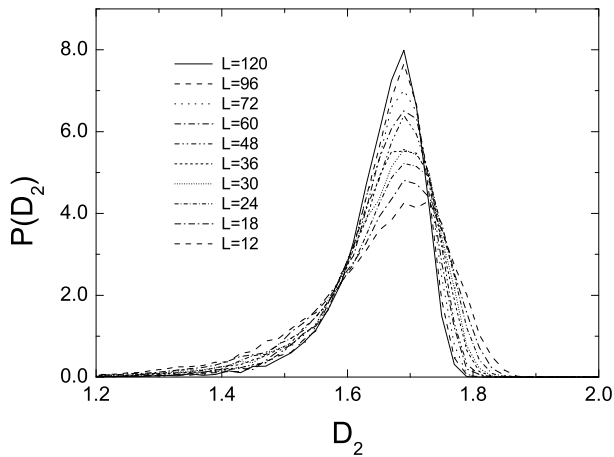}
\caption{Distribution functions of $D_2$ for the same ensembles
as for Fig.~3. Widths of the distribution functions become
narrow with increasing system size from $L=12$ (dashed line) to
$L=120$ (solid line). } \label{fig:6}
\end{figure}

Figure 6 shows the standard deviations $\sigma(L)$ for the
original ensembles and the {\it refined} ensembles (i.e., the
ensembles in which ALS are eliminated). Error bars of $\sigma$
estimated by the bootstrap method are smaller than their marks.
According to a standard scaling analysis that the $L$-dependence
of a statistical quantity at the critical point is scaled by an
irrelevant length, we can write $\sigma(L)$ as
\begin{equation}
\sigma(L)=\sigma_\infty + c L^{-y},
\label{eq:19}
\end{equation}
where $y$ is an irrelevant exponent and $\sigma_\infty$ is the
standard deviation for $L \rightarrow \infty$. Fitting data for
the original ensembles to Eq.~(\ref{eq:19}), we obtain
$\sigma_\infty=0.032\pm0.018$ and $y=0.39\pm0.12$. The value of
$\sigma_\infty$ is positive finite even though its error bar is
taken into account. More precisely, $\sigma_\infty$ is not zero
with the confidence coefficient $93\%$. This implies that $D_2$
fluctuates even in the thermodynamic limit as shown by Ref.~35. On
the contrary, fitting data for the refined ensembles gives
$\sigma_\infty=0.005\pm0.013$ and $y=0.42\pm0.08$. In this case,
$\sigma_\infty$ is very close to zero, and the error bar of
$\sigma_\infty$ is larger than its mean value. Therefore, it is
natural to deduce that $\sigma_\infty$ is zero. We can then
conclude that the correlation dimension $D_2$ for {\it typical}
critical wavefunctions does not fluctuate in the thermodynamic
limit. It should be noted that ALS were also eliminated in the
numerical work of Ref.~11 which supports the non-fluctuating $D_2$
for $L \rightarrow \infty$.

\begin{figure}[t]
\includegraphics[width=8cm]{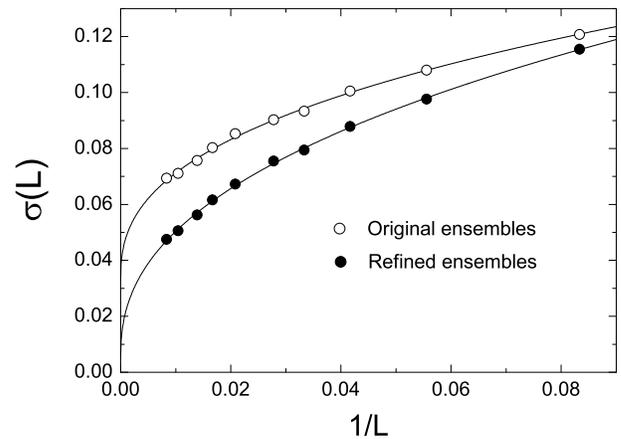}
\caption{ Standard deviations of the distribution functions
$P(D_2)$ as a function of $L^{-1}$. Open and filled circles
indicate results for the original ensembles and the refined
ensembles, respectively. Solid lines show the scaling fits by
Eq.~(\ref{eq:19}). } \label{fig:7}
\end{figure}

Finally, we consider the exponent $y$ in Eq.~(\ref{eq:19}). In the
SU(2) model, the scaling correction due to a spin-relaxation
length is quite small. What is the origin of the irrelevant
exponent $y$ in this case? We emphasize that the value of $y$ is
close to $d-D_2$ within the numerical error, where
$D_2=1.66\pm0.05$ estimated for the refined ensemble for $L=120$.
The relation $y=d-D_2$ has been analytically predicted by
Polyakov,\cite{Polyakov1} where $y$ is the exponent describing the
correction related to the quantum return probability. Thus, we
claim that the exponent $y$ describing the system-size dependence
of the standard deviation $\sigma(L)$ for the refined ensembles is
nothing but the irrelevant exponent predicted by Polyakov, which
is independent of microscopic details of systems.

\section{Conclusions}
\label{sec:6}

We have studied statistical properties of anomalously localized
states (ALS) at the Anderson transition point by defining ALS
quantitatively. In our definition, a wavefunction is regarded as
ALS when the functions $Q_L(l)$ and $R_L(r)$ which are the special
cases of the box-measure correlation functions $G_q(l,L,r)$ do not
follow the power laws Eqs.~(\ref{eq:6}) and (\ref{eq:7}).
Applying the definition of ALS to ensembles of critical
wavefunctions of the SU(2) model which describes a two-dimensional
electron system with strong spin-orbit interactions, it has been
revealed that the probability to find ALS at criticality increases
with the system size, while typical states are multifractal. This
result suggests that ALS should be eliminated from an ensemble of
critical wavefunctions if we study critical properties from
distributions of critical quantities. We also found that the
distribution function of $\Gamma$ is characterized only by its
average value $\bar \Gamma$, where $\Gamma$ quantifies
non-multifractality of a wavefunction. The influence of ALS to
critical properties in infinite systems has been demonstrated by
investigating the distribution of the correlation dimension $D_2$.
While the distribution function of $D_2$ has a finite width in the
thermodynamic limit if ALS are not eliminated from an ensemble of
critical wavefunctions, $D_2$ takes a definite value for $L
\rightarrow \infty$ if ALS are eliminated.

The existence of ALS at the critical point gives crucial
influences also for transport properties at the Anderson
transition point. The value of conductance for ALS is small, while
it is relatively large for multifractal states. The fluctuation
of critical conductance is deeply related to the ALS distribution
such as Fig.~3. Furthermore, the frequency dependence of ac
conductivity is affected by ALS as well, because strongly
localized ALS contribute only to high-frequency ac transport. Our
numerical works have been performed for the SU(2) model. We believe,
however, that properties of ALS are essentially the same for
systems belonging to other universality classes which exhibit the
Anderson transition. Further quantitative investigations of ALS
reveal concrete relations between the nature of ALS and physical
phenomena.

\begin{acknowledgements}
We are grateful to T. Nakayama, T. Ohtsuki and K. Slevin for
helpful discussions. This work was supported in part by a
Grant-in-Aid for Scientific Research from Japan Society for the
Promotion of Science (No.~$14540317$). Numerical calculations in
this work have been mainly performed on the facilities of the
Supercomputer Center,Institute for Solid State Physics, University
of Tokyo.
\end{acknowledgements}


\begin{thebibliography}{99}
\bibitem{Obuse1}
Details of the forced oscillator method for quaternion-real
matrices will be published elsewhere.
\end{thebibliography}
\end{document}